\pdfoutput=1
\documentclass[12pt,a4paper]{article}
\usepackage{ifthen} 
\newboolean{pdflatex}
\setboolean{pdflatex}{true} 

\usepackage{amsmath}
\usepackage{amssymb}
\usepackage{mathtools}

\newboolean{articletitles}
\setboolean{articletitles}{true} 

\newboolean{uprightparticles}
\setboolean{uprightparticles}{false} 

\newboolean{inbibliography}
\setboolean{inbibliography}{false} 

\usepackage{overpic} 
\usepackage{multirow} 
\usepackage{booktabs} 
\usepackage{tabularx} 

\usepackage{microtype}
\usepackage{lineno}  
\usepackage{xspace} 
\usepackage{caption} 

\usepackage{graphicx}  
\usepackage{color}
\usepackage{colortbl}
\graphicspath{{./figs/}} 

\usepackage{amsmath} 
\usepackage{amssymb}
\usepackage{amsfonts}
\usepackage{upgreek} 

\newcommand*\patchAmsMathEnvironmentForLineno[1]{%
\expandafter\let\csname old#1\expandafter\endcsname\csname #1\endcsname
\expandafter\let\csname oldend#1\expandafter\endcsname\csname
end#1\endcsname
 \renewenvironment{#1}%
   {\linenomath\csname old#1\endcsname}%
   {\csname oldend#1\endcsname\endlinenomath}%
}
\newcommand*\patchBothAmsMathEnvironmentsForLineno[1]{%
  \patchAmsMathEnvironmentForLineno{#1}%
  \patchAmsMathEnvironmentForLineno{#1*}%
}
\AtBeginDocument{%
\patchBothAmsMathEnvironmentsForLineno{equation}%
\patchBothAmsMathEnvironmentsForLineno{align}%
\patchBothAmsMathEnvironmentsForLineno{flalign}%
\patchBothAmsMathEnvironmentsForLineno{alignat}%
\patchBothAmsMathEnvironmentsForLineno{gather}%
\patchBothAmsMathEnvironmentsForLineno{multline}%
}

\usepackage{hyperref}    
\usepackage[all]{hypcap} 

\def\lhcb {\mbox{LHCb}\xspace}

\ifthenelse{\boolean{uprightparticles}}%
{

 \def\Pmu         {\ensuremath{\upmu}\xspace}

 \def\Ppsi        {\ensuremath{\uppsi}\xspace}

 \def\PDelta      {\ensuremath{\Delta}\xspace}                 
 \def\PXi      {\ensuremath{\Xi}\xspace}                 
 \def\PLambda      {\ensuremath{\Lambda}\xspace}                 
 \def\PSigma      {\ensuremath{\Sigma}\xspace}                 
 \def\POmega      {\ensuremath{\Omega}\xspace}                 
 \def\PUpsilon      {\ensuremath{\Upsilon}\xspace}                 
 
 \def\PB      {\ensuremath{\mathrm{B}}\xspace}                 
                  
 \def\PD      {\ensuremath{\mathrm{D}}\xspace}

 \def\PJ      {\ensuremath{\mathrm{J}}\xspace}                 
 \def\PK      {\ensuremath{\mathrm{K}}\xspace}

 \def\PZ      {\ensuremath{\mathrm{Z}}\xspace}                 
                  
 \def\Pb      {\ensuremath{\mathrm{b}}\xspace}                 
 \def\Pc      {\ensuremath{\mathrm{c}}\xspace}

 \def\Pi      {\ensuremath{\mathrm{i}}\xspace}

}
{

 \def\Pmu         {\ensuremath{\mu}\xspace}

 \def\Ppsi        {\ensuremath{\psi}\xspace}                 
                  
 \mathchardef\PDelta="7101
 \mathchardef\PXi="7104
 \mathchardef\PLambda="7103
 \mathchardef\PSigma="7106
 \mathchardef\POmega="710A
 \mathchardef\PUpsilon="7107
                  
 \def\PB      {\ensuremath{B}\xspace}                 
                  
 \def\PD      {\ensuremath{D}\xspace}

 \def\PJ      {\ensuremath{J}\xspace}                 
 \def\PK      {\ensuremath{K}\xspace}

 \def\PZ      {\ensuremath{Z}\xspace}                 
                  
 \def\Pb      {\ensuremath{b}\xspace}                 
 \def\Pc      {\ensuremath{c}\xspace}

 \def\Pi      {\ensuremath{i}\xspace}

}

\def\mumu       {{\ensuremath{\Pmu^+\Pmu^-}}\xspace}

\def\Z      {{\ensuremath{\PZ}}\xspace}

\def\cquark    {{\ensuremath{\Pc}}\xspace}
\def\cquarkbar {{\ensuremath{\overline \cquark}}\xspace}
\def\ccbar     {{\ensuremath{\cquark\cquarkbar}}\xspace}
\def\bquark    {{\ensuremath{\Pb}}\xspace}
\def\bquarkbar {{\ensuremath{\overline \bquark}}\xspace}
\def\bbbar     {{\ensuremath{\bquark\bquarkbar}}\xspace}

  \def\Kbar    {{\kern 0.2em\overline{\kern -0.2em \PK}{}}\xspace}

  \def\Dbar    {{\kern 0.2em\overline{\kern -0.2em \PD}{}}\xspace}

\def\Bbar    {{\ensuremath{\kern 0.18em\overline{\kern -0.18em \PB}{}}}\xspace}

\def\jpsi     {{\ensuremath{{\PJ\mskip -3mu/\mskip -2mu\Ppsi\mskip 2mu}}}\xspace}

  \def\Y#1S{\ensuremath{\PUpsilon{(#1S)}}\xspace}

\def\Lbar        {{\ensuremath{\kern 0.1em\overline{\kern -0.1em\PLambda}}}\xspace}

\newcommand{\decay}[2]{\ensuremath{#1\!\to #2}\xspace}         

\def\to                 {\ensuremath{\rightarrow}\xspace}

\def\AT#1     {\ensuremath{A_{\mathrm{T}}^{#1}}\xspace}           

\def\C#1      {\ensuremath{\mathcal{C}_{#1}}\xspace}                       
\def\Cp#1     {\ensuremath{\mathcal{C}_{#1}^{'}}\xspace}                    
\def\Ceff#1   {\ensuremath{\mathcal{C}_{#1}^{\mathrm{(eff)}}}\xspace}        
\def\Cpeff#1  {\ensuremath{\mathcal{C}_{#1}^{'\mathrm{(eff)}}}\xspace}       
\def\Ope#1    {\ensuremath{\mathcal{O}_{#1}}\xspace}                       
\def\Opep#1   {\ensuremath{\mathcal{O}_{#1}^{'}}\xspace}                    

\newcommand{\tev}{\ifthenelse{\boolean{inbibliography}}{\ensuremath{~T\kern -0.05em eV}\xspace}{\ensuremath{\mathrm{\,Te\kern -0.1em V}}}\xspace}
\newcommand{\gev}{\ensuremath{\mathrm{\,Ge\kern -0.1em V}}\xspace}
\newcommand{\mev}{\ensuremath{\mathrm{\,Me\kern -0.1em V}}\xspace}
\newcommand{\kev}{\ensuremath{\mathrm{\,ke\kern -0.1em V}}\xspace}
\newcommand{\ev}{\ensuremath{\mathrm{\,e\kern -0.1em V}}\xspace}
\newcommand{\gevc}{\ensuremath{{\mathrm{\,Ge\kern -0.1em V\!/}c}}\xspace}
\newcommand{\mevc}{\ensuremath{{\mathrm{\,Me\kern -0.1em V\!/}c}}\xspace}
\newcommand{\gevcc}{\ensuremath{{\mathrm{\,Ge\kern -0.1em V\!/}c^2}}\xspace}
\newcommand{\gevgevcccc}{\ensuremath{{\mathrm{\,Ge\kern -0.1em V^2\!/}c^4}}\xspace}
\newcommand{\mevcc}{\ensuremath{{\mathrm{\,Me\kern -0.1em V\!/}c^2}}\xspace}

\def\mum  {\ensuremath{{\,\upmu\rm m}}\xspace}

\def\invnb {\ensuremath{\mbox{\,nb}^{-1}}\xspace}

\def\gsim{{~\raise.15em\hbox{$>$}\kern-.85em
          \lower.35em\hbox{$\sim$}~}\xspace}
\def\lsim{{~\raise.15em\hbox{$<$}\kern-.85em
          \lower.35em\hbox{$\sim$}~}\xspace}

\def\pt         {\mbox{$p_{\rm T}$}\xspace}

\def\evtgen     {\mbox{\textsc{EvtGen}}\xspace}

\def\geant      {\mbox{\textsc{Geant4}}\xspace}

\def\photos     {\mbox{\textsc{Photos}}\xspace}

\def\pythia     {\mbox{\textsc{Pythia6}}\xspace}

\def\tell1  {TELL1\xspace}
\def\ukl1   {UKL1\xspace}

\newcommand{\ie}{\mbox{\itshape i.e.}\xspace}

\usepackage{cite} 
\usepackage{mciteplus}

\usepackage{longtable} 

\begin{document}

\renewcommand{\thefootnote}{\fnsymbol{footnote}}
\setcounter{footnote}{1}


\begin{titlepage}
\pagenumbering{roman}

\vspace*{-1.5cm}
\centerline{\large EUROPEAN ORGANIZATION FOR NUCLEAR RESEARCH (CERN)}
\vspace*{1.5cm}
\hspace*{-0.5cm}
\begin{tabular*}{\linewidth}{lc@{\extracolsep{\fill}}r}
\ifthenelse{\boolean{pdflatex}}
{\vspace*{-2.7cm}\mbox{\!\!\!\includegraphics[width=.14\textwidth]{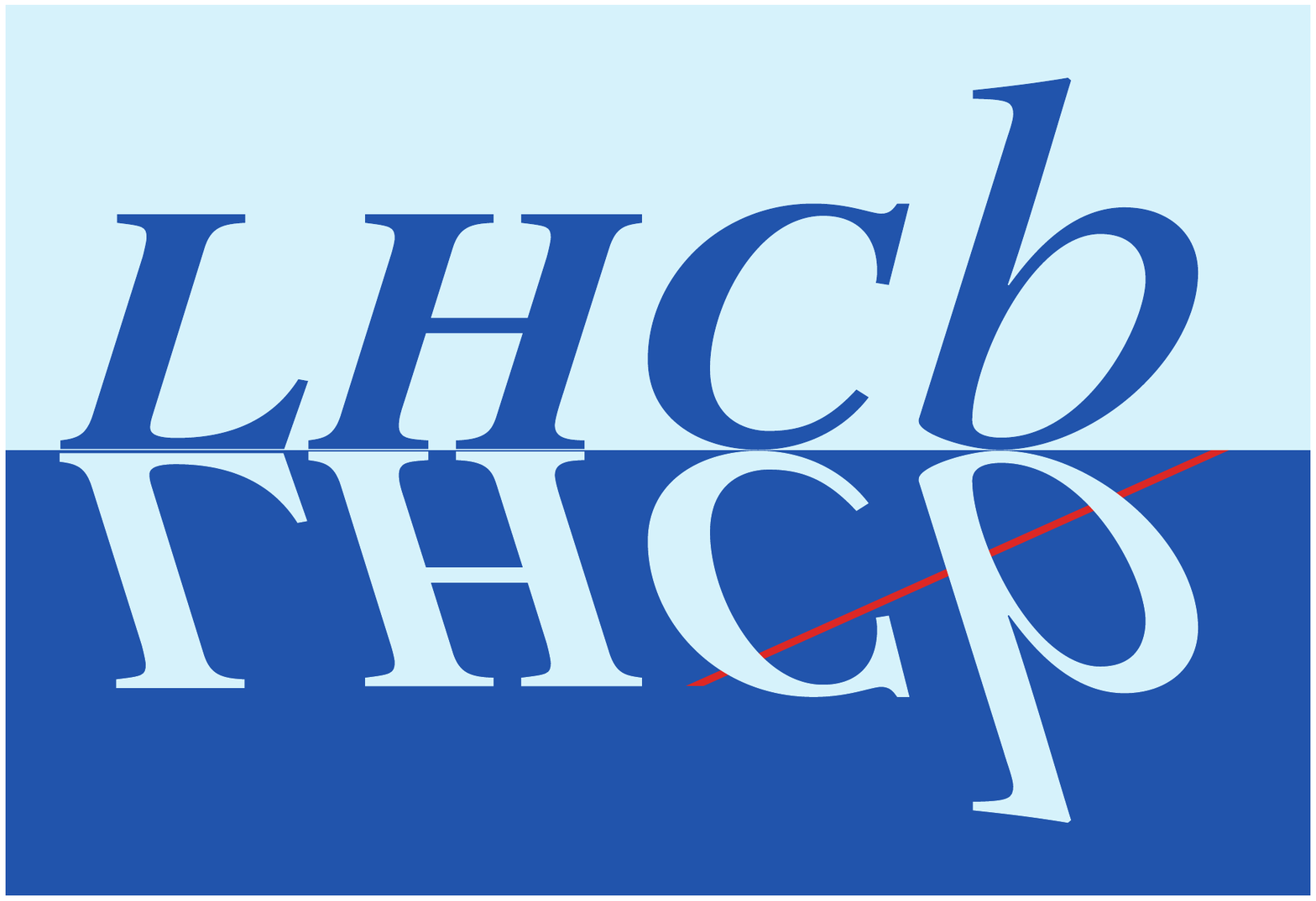}} & &}%
{\vspace*{-1.2cm}\mbox{\!\!\!\includegraphics[width=.12\textwidth]{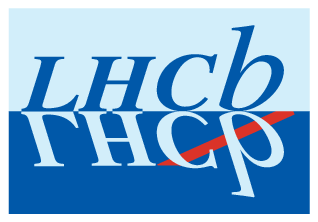}} & &}%
\\
 & & CERN-PH-EP-2014-126 \\  
 & & LHCb-PAPER-2014-022 \\  
 & & \today \\ 
 & & \\
\end{tabular*}

\vspace*{1.0cm}

{\bf\boldmath\huge
\begin{center}
  Observation of \Z production in proton-lead collisions at LHCb
\end{center}
}

\vspace*{1.5cm}

\begin{center}
The LHCb collaboration\footnote{Authors are listed on the following pages.}
\end{center}

\vspace{\fill}

\begin{abstract}
  \noindent
  The first observation of \Z boson production in proton-lead collisions at a centre-of-mass energy per proton-nucleon pair of $\sqrt{s_{NN}}=5\tev$ is presented. The data sample corresponds to an integrated luminosity of 1.6\invnb collected with the LHCb detector. The \Z candidates are reconstructed from pairs of oppositely charged muons with pseudorapidities between 2.0 and 4.5 and transverse momenta above 20\gevc. The invariant dimuon mass is restricted to the range $60-120\gevcc$. The \Z production cross-section is measured to be
  \begin{align*}
	\sigma_{\Z\to\mumu}(\text{fwd})&=\;13.5^{+5.4}_{-4.0}\text{(stat.)}\pm1.2\text{(syst.)}~\text{nb}\\
	\intertext{in the direction of the proton beam and}
	\sigma_{\Z\to\mumu}(\text{bwd}) & =\;10.7^{+8.4}_{-5.1}\text{(stat.)}\pm1.0\text{(syst.)}~\text{nb}
  \end{align*}
in the direction of the lead beam, where the first uncertainty is statistical and the second systematic.
  
\end{abstract}

\vspace*{1.0cm}

\begin{center}
  Published in JHEP 09 (2014) 030 
\end{center}

\vspace{\fill}

{\footnotesize 
\centerline{\copyright~CERN on behalf of the \lhcb collaboration, license \href{http://creativecommons.org/licenses/by/3.0/}{CC-BY-3.0}.}}
\vspace*{2mm}

\end{titlepage}


\newpage
\setcounter{page}{2}
\mbox{~}
\newpage

\centerline{\large\bf LHCb collaboration}
\begin{flushleft}
\small
R.~Aaij$^{41}$, 
B.~Adeva$^{37}$, 
M.~Adinolfi$^{46}$, 
A.~Affolder$^{52}$, 
Z.~Ajaltouni$^{5}$, 
S.~Akar$^{6}$, 
J.~Albrecht$^{9}$, 
F.~Alessio$^{38}$, 
M.~Alexander$^{51}$, 
S.~Ali$^{41}$
% ...

{\footnotesize \it
% ...
$ ^{5}$Clermont Universit\'{e}, Universit\'{e} Blaise Pascal, CNRS/IN2P3, LPC, Clermont-Ferrand, France\\
$ ^{6}$CPPM, Aix-Marseille Universit\'{e}, CNRS/IN2P3, Marseille, France\\
% ...
$ ^{9}$Fakult\"{a}t Physik, Technische Universit\"{a}t Dortmund, Dortmund, Germany\\
% ...
$ ^{37}$Universidad de Santiago de Compostela, Santiago de Compostela, Spain\\
$ ^{38}$European Organization for Nuclear Research (CERN), Geneva, Switzerland\\
% ...
$ ^{41}$Nikhef National Institute for Subatomic Physics, Amsterdam, The Netherlands\\
% ...
$ ^{46}$H.H. Wills Physics Laboratory, University of Bristol, Bristol, United Kingdom\\
% ...
$ ^{51}$School of Physics and Astronomy, University of Glasgow, Glasgow, United Kingdom\\
$ ^{52}$Oliver Lodge Laboratory, University of Liverpool, Liverpool, United Kingdom\\
% ...
}
\end{flushleft}

\cleardoublepage


\renewcommand{\thefootnote}{\arabic{footnote}}
\setcounter{footnote}{0}



\pagestyle{plain} 
\setcounter{page}{1}
\pagenumbering{arabic}

\section{Introduction\label{sec:Intro}}
Measurements in proton-nucleus collisions can serve as references for nucleus-nucleus collisions and be used as inputs for the determination of nuclear parton distribution functions (nPDF)~\cite{Salgado:2012}. The nPDF, $f_a^{i,A}(x_A,Q^2)$, is defined as a function of the momentum fraction, $x_A$, of a certain parton type $a$ inside the nucleon $i$ bound in a nucleus $A$ for an energy scale $Q^2$. It is usually parametrised by the nuclear modification factor, $R_a^A(x_A,Q^2)$, which represents the ratio between the nPDF and the corresponding free nucleon baseline PDF, $f_a^i(x,Q^2)$. For $x_A$ less than 0.02, $R_a^A$ is smaller than unity as the coherence length of the interaction is larger than the nucleon-nucleon distance inside the nucleus (shadowing effect)~\cite{Glauber:1955}. For $x_A$-values between about 0.4 and 0.8, $R_a^A$ is less than unity  due to the EMC effect~\cite{Aubert:1983}. For values of $x_A$ close to 1.0, $R_a^A$ is predicted to be greater than unity due to Fermi motion~\cite{Eskola:2013,*Paukkunen:2014}. In the intermediate region between the regimes of shadowing and the EMC effect, $R_a^A$ is also predicted to be greater than unity because of the sum rule of PDFs (anti-shadowing)~\cite{Glauber:1955a,Kharzeev:1994,Frankfurt:2012}.\\
\indent The most recent available nPDF sets~\cite{Eskola:2009,Hirai:2007,*Hirai:2007a,Florian:2012,Schienbein:2009,*Kovarik:2011,*Kovarik:2013} are primarily based on data from fixed target experiments (deep inelastic scattering (DIS) and Drell-Yan processes) and, in particular cases, on additional data from d-Au collisions at RHIC and from neutrino DIS. As a consequence of the kinematic range in these experiments, nPDFs for $Q^2$ larger than 10~GeV$^2$ have either no or only weak direct constraints for values of $x_A$ smaller than 0.01 and $x_A$ close to one~\cite{Salgado:2012,Eskola:2013,*Paukkunen:2014}.\\
\indent Since measurements of electroweak boson production in the forward direction of proton-lead collisions involve small $x_A$-values and in the backward direction $x_A$-values close to one, they can serve as input to nPDF fits with a significant constraining power, especially at small $x_A$-values. `Forward' refers to positive rapidity values defined relative to the direction of the proton beam, `backward' to negative rapidity values. No measurement of electroweak boson production in proton-lead collisions has yet been performed, while their production in lead-lead collisions has been measured for central rapidities~\cite{Aad:2011,*ATLAS:2011,Chatrchyan:2011,*Chatrchyan:2012b}.\\
\indent Due to its rapidity coverage, the LHCb experiment has the ability to probe $x_A$ down to very small, but also up to very large values. The sensitivity of $x_A$ at a centre-of-mass energy per proton-nucleon pair of $\sqrt{s_{NN}}=5~\text{TeV}$ varies from $2\times10^{-4}$ to $3\times10^{-3}$ in the forward case and from 0.2 to 1.0 in the backward case at an energy scale $Q^2=M_\Z^2$.\\
\indent In this paper a first measurement of the \Z production cross-section in proton-lead collisions in the forward and backward region is presented. The measurement uses \Z bosons reconstructed in the \mumu final state and is performed in a similar manner to the \Z production measurements in $pp$ collisions described in Refs.~\cite{LHCb-PAPER-2012-008} and \cite{LHCb-CONF-2013-007}. The fiducial region in the laboratory frame is defined by: the transverse momenta, \pt, of the muons, which must be larger than 20\gevc; their pseudorapidities, $\eta$, required to be between 2.0 and 4.5; and the invariant dimuon mass, which is restricted to be between 60 and 120\gevcc. The cross-section is measured as
\begin{equation}
\sigma_{\decay{\Z}{\mumu}} = \frac{N_\text{cand}\times\rho}{\mathcal{L}\times\varepsilon},
\label{eq:CrossSec}
\end{equation}
where $N_\text{cand}$ is the number of selected \Z candidates, $\rho$ the purity of the sample, $\mathcal{L}$ the integrated luminosity and $\varepsilon$ the total efficiency.\\
\indent The results are compared to predictions at next-to-next-to-leading order (NNLO~\cite{Gavin:2011,*Li:2012}) in perturbative quantum chromodynamics with and without nuclear effects~\cite{Eskola:2009}.

\section{Detector and software\label{sec:Detector}}
The \lhcb detector~\cite{Alves:2008zz} is a single-arm forward
spectrometer covering the \mbox{pseudorapidity} range $2<\eta <5$,
designed for the study of particles containing \bquark or \cquark
quarks. The detector includes a high-precision tracking system
consisting of a silicon-strip vertex detector surrounding the
interaction region, a large-area silicon-strip detector located
upstream of a dipole magnet with a bending power of about
$4{\rm\,Tm}$, and three stations of silicon-strip detectors and straw
drift tubes~\cite{LHCb-DP-2013-003} placed downstream.
The combined tracking system provides a momentum measurement with
relative uncertainty that varies from 0.4\,\% at 5\gevc to 0.6\,\% at 100\gevc,
and an impact parameter resolution of 20\mum for
tracks with large transverse momentum. The energy of photons, electrons and hadrons is measured by a calorimeter system consisting of
scintillating-pad and preshower detectors, an electromagnetic
calorimeter (ECAL) and a hadronic calorimeter (HCAL). Muons are identified by a
system composed of alternating layers of iron and multiwire
proportional chambers~\cite{LHCb-DP-2012-002}.
The trigger~\cite{LHCb-DP-2012-004} consists of a
hardware stage, based on information from the calorimeter and muon
systems, followed by a software stage, which applies a full event
reconstruction.\\
\indent During data taking of proton-lead collisions, the hardware stage of the trigger  accepted all non-empty bunch crossings, and the software stage accepted events with at least one well-reconstructed track matched to hits in the muon system with a momentum above 8\gevc and \pt above 4.8\gevc. In order to avoid events with high hit multiplicity, which dominate the processing time in the software trigger, global event cuts (GEC) are applied on hit multiplicities in the tracking detectors.\\
\indent Simulated $pp$ collisions used in  the estimation of the purity are generated using
\pythia~\cite{Sjostrand:2006za} with a specific \lhcb
configuration~\cite{LHCb-PROC-2010-056}. Decays of hadronic particles
are described by \evtgen~\cite{Lange:2001uf}, in which final state
radiation is generated using \photos~\cite{Golonka:2005pn}. The
interaction of the generated particles with the detector and its
response are implemented using the \geant toolkit~\cite{Allison:2006ve, *Agostinelli:2002hh} as described in
Ref.~\cite{LHCb-PROC-2011-006}.\\
\indent To study the kinematics in proton-lead collisions, dedicated samples of simulated events at parton level are generated with \textsc{Pythia8}~\cite{Sjostrand:2007gs} together with the MSTW08 PDF set~\cite{MSTW08LO}. These samples contain \decay{\Z}{\mumu} decays produced in proton-proton and proton-neutron collisions  at a centre-of-mass energy of $\sqrt{s}=5\tev$ with beam energies corresponding to the energy of the proton beam and the energy per nucleon in the lead beam, respectively.
\section{Data samples and candidate selection\label{sec:Sel}}
\begin{figure}[b]
\begin{center}
\begin{minipage}[t]{0.49\textwidth}
\begin{overpic}[width=\textwidth,scale=.25,tics=20]{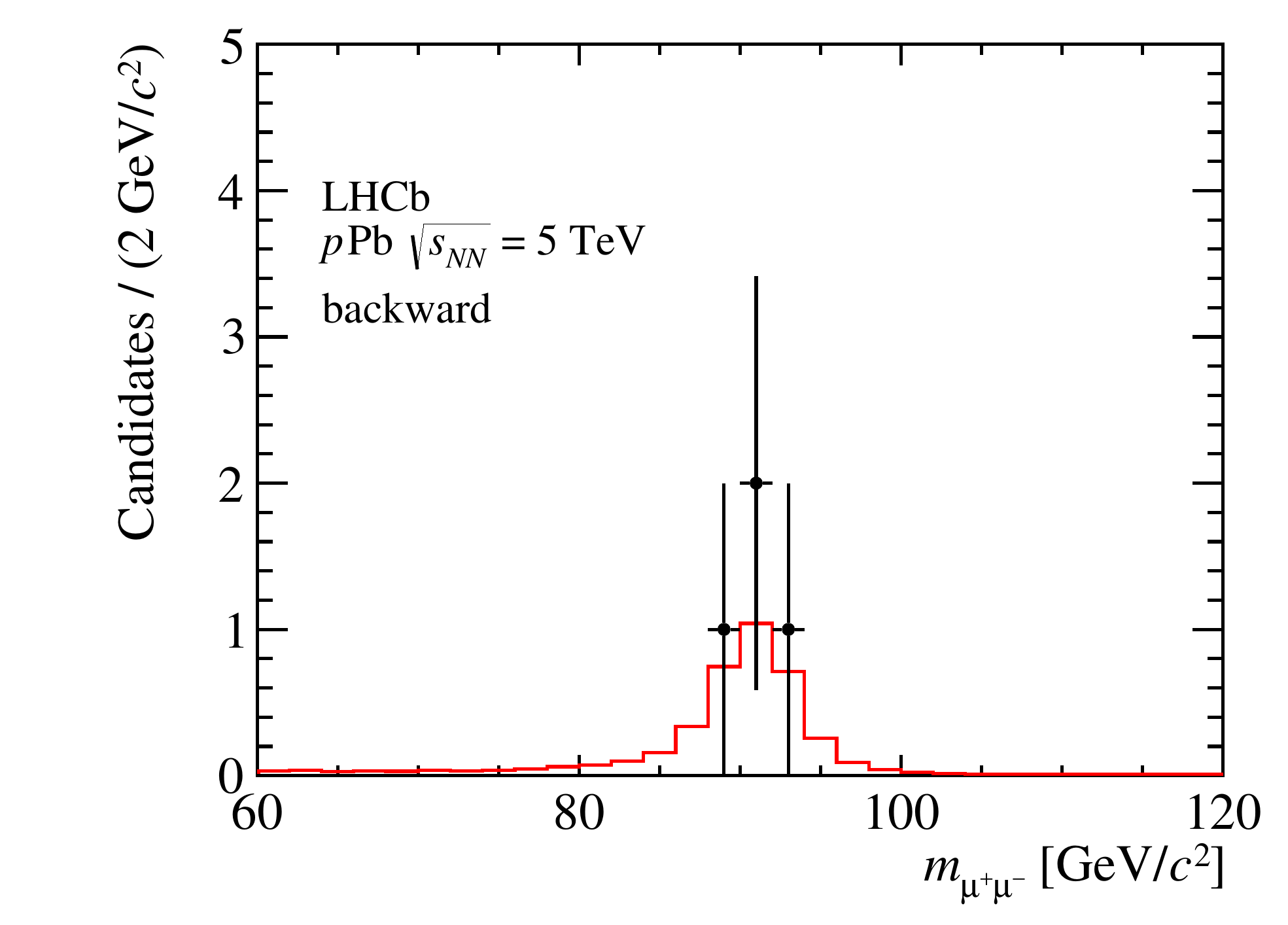}
\put(25,30){(a)}
\put(62,45){\includegraphics[width=0.32\textwidth]{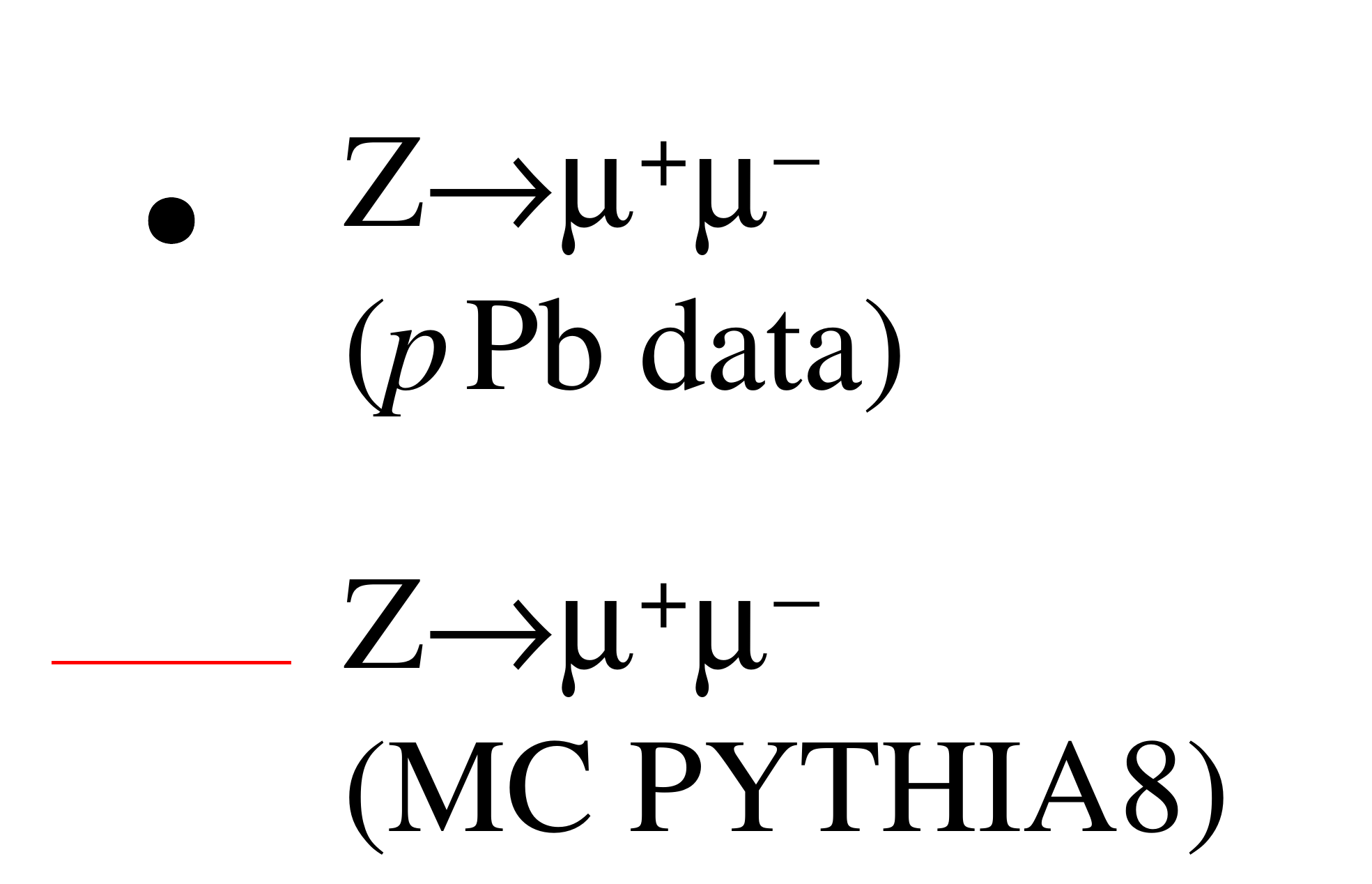}}
\end{overpic}
\end{minipage}
\begin{minipage}[t]{0.49\textwidth}
\begin{overpic}[width=\textwidth,scale=.25,tics=20]{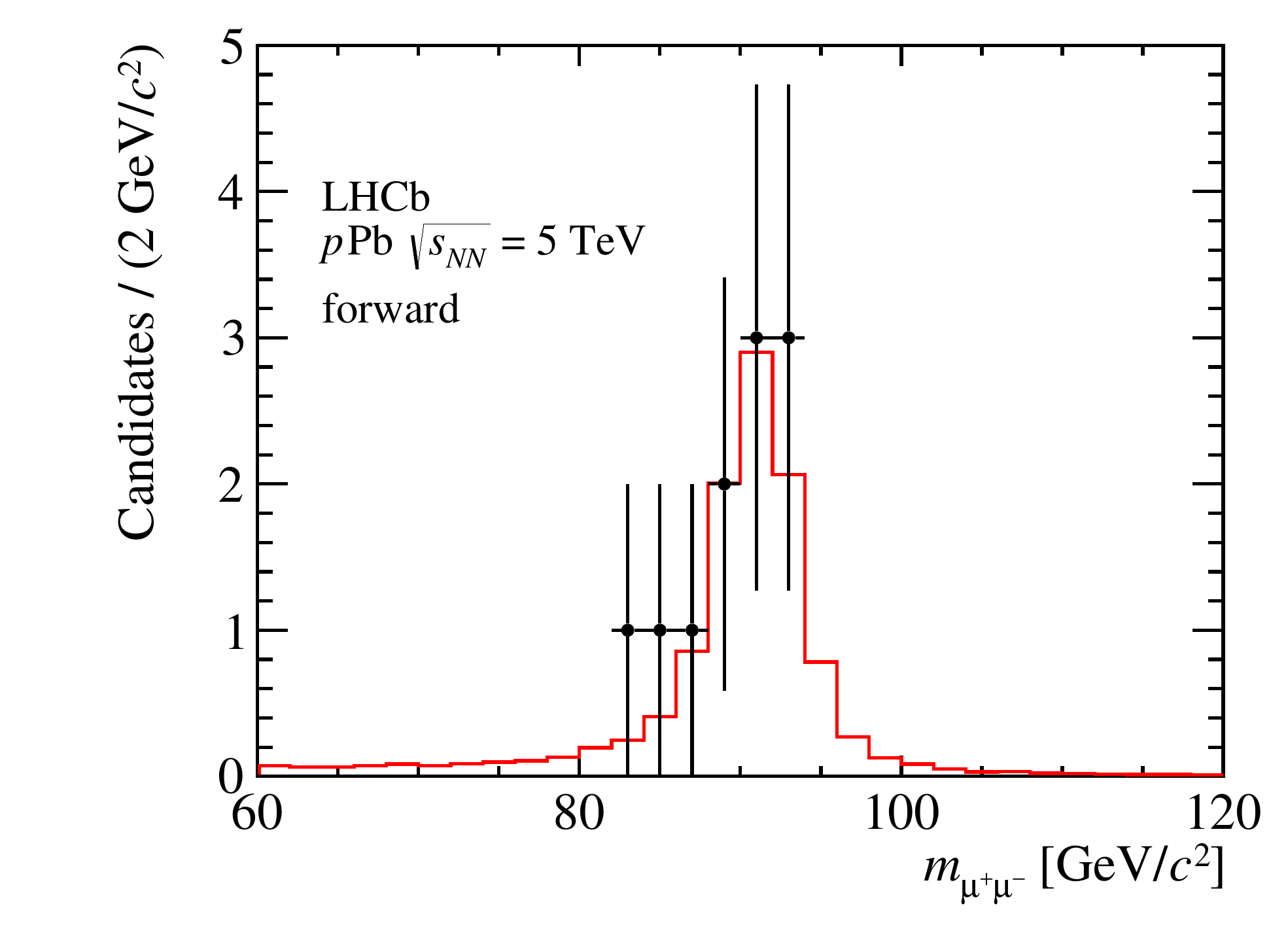}
\put(25,30){(b)}
\put(62,45){\includegraphics[width=0.32\textwidth]{leg1.pdf}}
\end{overpic}\
\end{minipage}
\caption[Invariant dimuon mass]{Invariant dimuon mass distribution of selected \Z candidates in (a) the backward and (b) the forward sample are shown by the black data points with error bars. The red line shows the distribution obtained from simulation using \textsc{Pythia8} with the MSTW08 PDF set normalised to the number of observed candidates.}
\label{fig:ZMass}
\end{center}
\end{figure}

\begin{figure}[bt]
\begin{center}
\begin{minipage}[t]{0.49\textwidth}
\begin{overpic}[width=\textwidth,scale=.25,tics=20]{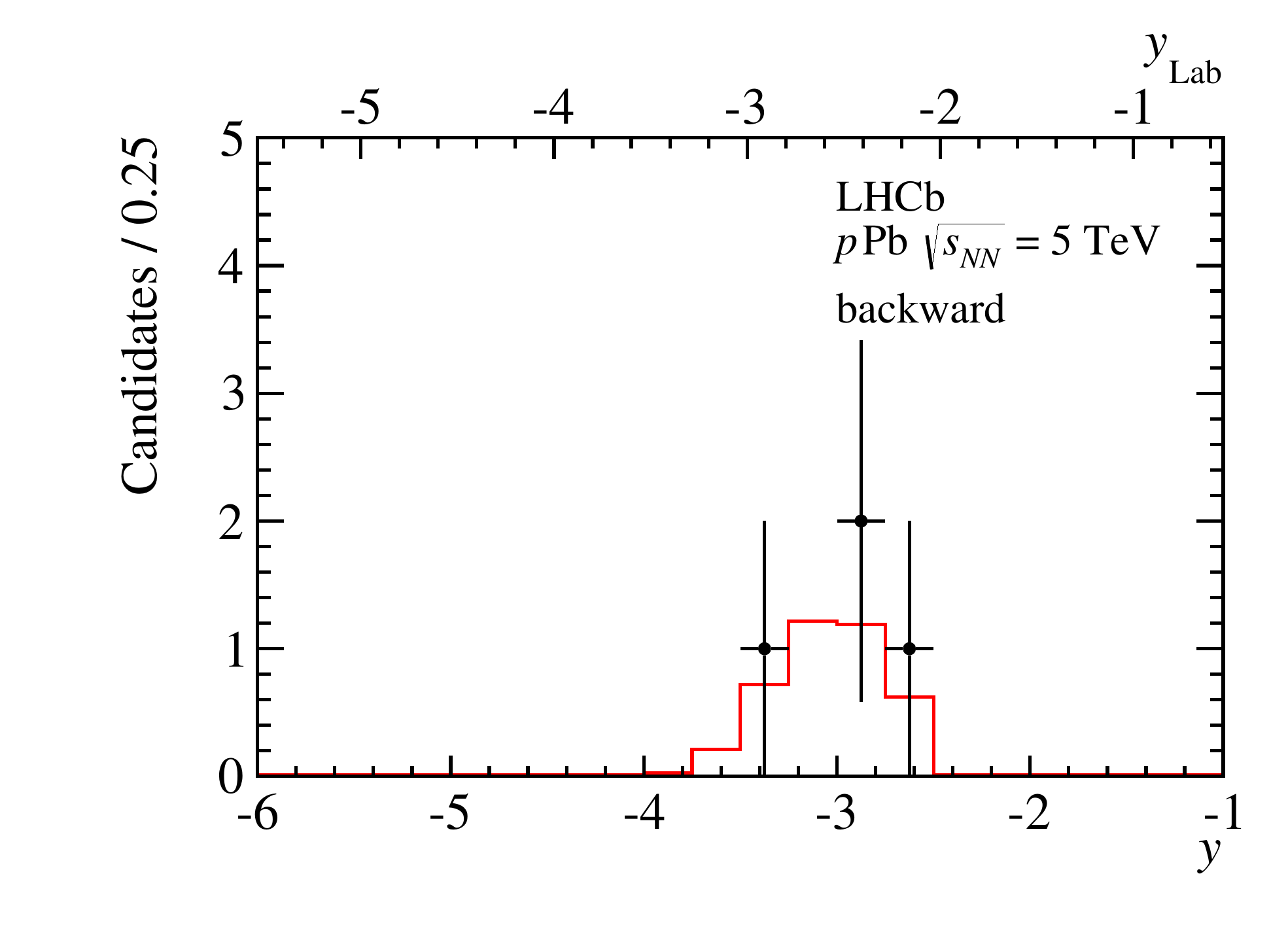}
\put(24,25){\includegraphics[width=0.32\textwidth]{leg1.pdf}}
\put(25,50){(a)}
\end{overpic}
\end{minipage}
\begin{minipage}[t]{0.49\textwidth}
\begin{overpic}[width=\textwidth,scale=.25,tics=20]{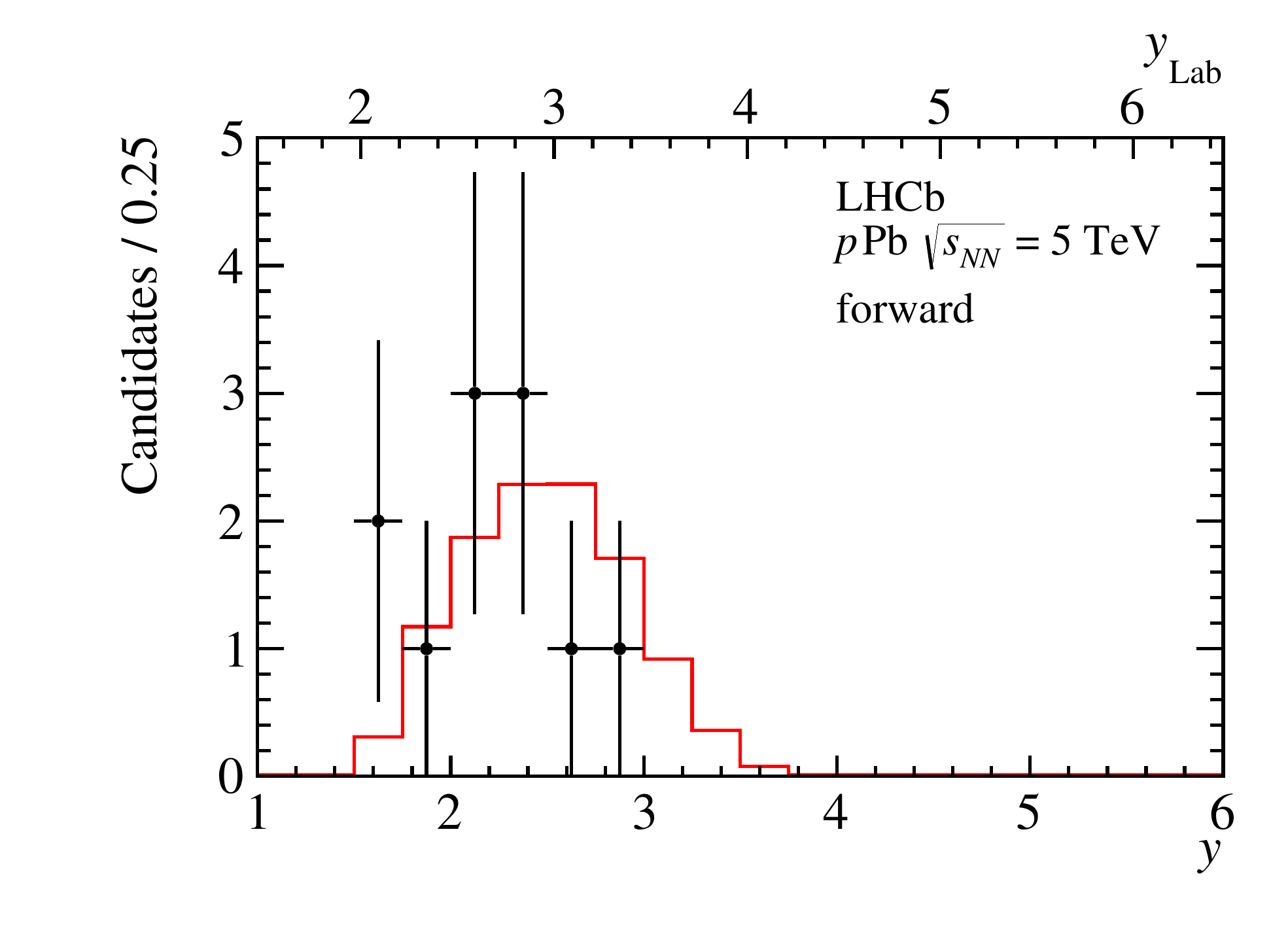}
\put(60,25){\includegraphics[width=0.32\textwidth]{leg1.pdf}}
\put(25,50){(b)}
\end{overpic}
\end{minipage}
\caption[Nuclear PDF]{Rapidity distribution of selected \Z candidates in (a) the backward and in (b) the forward sample are shown by the black data points. The red line shows the distribution obtained from simulation using \textsc{Pythia8} with the MSTW08 PDF set normalised to the number of observed candidates. The top $x$-axis shows the rapidity, $y_\text{Lab}$, in the laboratory frame, the bottom one the rapidity, $y$, in the centre-of-mass frame.}
\label{fig:Zrap}
\end{center}
\end{figure}

The analysis is based on data samples of proton-lead collisions that correspond to integrated luminosities of $1.099\pm0.021~\text{nb$^{-1}$}$ in the forward and $0.521\pm0.011~\text{nb$^{-1}$}$ in the backward direction. The integrated luminosity has been calibrated by Van der Meer scans separately for each beam configuration~\cite{Meer:1968,Burkhardt:2007}. The energy of the proton beam is $E_p = 3988\pm26$~GeV while the energy of the lead beam per nucleon is $E_N=1572\pm10$~GeV. This gives a centre-of-mass energy per proton-nucleon pair of $\sqrt{s_{NN}}=5008\pm33$~GeV, which is approximated to 5\tev~\cite{Wenninger:2013}. Due to the asymmetric beam energy there is a rapidity shift between the rapidity $y_\text{Lab}$ in the laboratory and $y$ in the centre-of-mass frame of $\Delta y = y_\text{lab} - y = +0.47$.\\
\indent The \Z candidates are selected from reconstructed pairs of oppositely charged particles that are identified as muons by matched hits in the muon system and that are inside the considered fiducial region. Selection criteria are applied to reject background candidates that fake \decay{\Z}{\mumu} decays: the impact parameter of each track with respect to the closest primary vertex must be less than 100~$\upmu$m, and the sum of the energies measured in the ECAL and HCAL associated to each track has to be less than 0.5 times the track momentum measured in the tracking system. The $\chi^2$-probability of the track fit is required to be larger than 1\,\% for both tracks.\\
\indent Figure~\ref{fig:ZMass} shows the invariant mass distribution of the selected candidates for the two beam configurations together with the predictions from simulation. There is a good agreement between data and simulation. In total, eleven candidates are selected in the forward sample and four in the backward sample. The rapidity distributions of the \Z candidates are shown in Fig.~\ref{fig:Zrap}.\\
\indent The efficiencies and the purity are studied with a sample of \Z candidates from $pp$ collisions at $\sqrt{s}=8\tev$ corresponding to 2~fb$^{-1}$ of integrated luminosity. These \Z candidates have to fulfil the same selection criteria as those in proton-lead collisions. Since the purity and efficiencies depend on the multiplicity, the difference in track multiplicity is taken into account by reweighting the candidates from the $pp$ sample to match the multiplicity in proton-lead collisions. The reweighting is performed using the observed ratio of the track multiplicity distributions of events containing \jpsi candidates in proton-lead and proton-proton collisions.

\section{Purity and efficiency determination\label{sec:Pureff}}
\indent The purity estimation considers two background sources. The first comprises candidates where at least one of the muons is a misidentified hadron. This background source is expected to have the same absolute abundance in oppositely charged as identically charged dimuon combinations and the relative fraction is estimated from the number of candidates observed in a sample of same-sign candidates and amounts to about 0.16\,\%.\\
\indent The second source, referred to as heavy-quark background, originates from semileptonic decays of \bbbar and \ccbar pairs into oppositely charged muon pairs. Two background-enriched samples are used. In the first sample, the vertex fit $\chi^2$ per degree of freedom of the dimuon vertex is required to be larger than 70 to reject signal combinations. In the second sample, an anti-isolation criterion on each muon rejects signal combinations: the \pt of the muon has to be less than 70\,\% of the summed \pt of of the tracks that differ in $\eta$ and the azimuthal angle, $\phi$, with respect to the muon track
by less than $\sqrt{\Delta\eta^2+\Delta\phi^2}=0.5$. In both cases, the efficiency of the cuts for the heavy-quark background is estimated from simulation. The relative background contribution is estimated from the efficiency-corrected background yield in the signal mass window. The two samples yield consistent results of about 0.20\,\%.\\
\indent The weighted average of the two estimates for heavy-quark background is added to the contribution from misidentified hadrons, which gives a purity, $\rho$, of $(99.74\pm0.06)\%$ for the forward and $(99.63\pm0.05)\%$ for the backward direction.\\
\indent The total efficiency $\varepsilon$ factorises into two parts as $\varepsilon = \varepsilon^\text{GEC}\times\varepsilon^\text{cand}$. The first term, $\varepsilon^\text{GEC}$, is related to the GEC, based on the occupancy in the vertex and tracking detectors. The second term, $\varepsilon^\text{cand}$, accounts for the reconstruction, selection, trigger and muon-identification efficiency of the \decay{\Z}{\mumu} decays.\\
\indent The value of $\varepsilon^\text{GEC}$ is estimated from the occupancy distributions in events with \jpsi candidates from proton-lead collisions following the method described in Ref.~\cite{LHCB-PAPER-2013-052}. While the GEC do not remove events in the forward beam configuration, they retain $(97.8\pm1.9)\%$ in the backward configuration.\\
\indent The reconstruction, selection, trigger and muon-identification efficiencies are estimated by tag-and-probe methods as described in Ref.~\cite{LHCb-PAPER-2012-008}. The tag-and-probe methods use track-multiplicity-reweighted \Z candidates from $pp$ collisions at $\sqrt{s}=8\tev$. These efficiencies are determined as functions of the muon pseudorapidity. As the sample size of the \Z candidates in proton-lead collisions is very small, the overall value of $\varepsilon^\text{cand}$ is evaluated by folding the efficiency as a function of the pseudorapidity with the $\eta$ distribution of muons. These distributions are obtained from simulated \Z bosons produced in $pp$ and $pn$ collisions to mimic proton-lead collisions. The value of $\varepsilon^\text{cand}$ is $(74.1\pm6.2)\%$ for the forward and $(72.8\pm6.3)\%$ for the backward direction.
\section{Systematic uncertainties\label{sec:Sys}}
Five potential sources of systematic uncertainties are considered:
\begin{enumerate}
\item The uncertainty on the sample purity is due to the statistical uncertainty on the data samples used in its determination. It also includes the statistical uncertainty on the efficiency determination of the cuts to obtain the samples enriched by the heavy-quark background using simulation. A further contribution is included for potential differences in the centre-of-mass energy dependence of the signal and background processes.
\item The uncertainty on the GEC efficiency is estimated by changing the functional form used to describe the occupancy distributions. Possible differences in the occupancy distributions between events containing \jpsi and those containing \Z candidates using data from $pp$ collisions at $\sqrt{s}=8\tev$ are taken into account.
\item The uncertainty on the candidate efficiency includes the uncertainties on the reconstruction, selection, trigger and muon-identification efficiencies. It is based on the statistical uncertainty of the measured efficiencies as well as on the uncertainty of the muon pseudorapidity spectrum from simulation used to obtain the average efficiency values.
\item The uncertainty on the track multiplicity reweighting is assigned as the relative difference in the ratio $\rho/\varepsilon^\text{cand}$ with and without applying the reweighting in the determination of the purity as well as the reconstruction, selection, trigger and muon-identification efficiencies.
\item The uncertainty on the luminosity is based on the statistical and systematic uncertainties of the calibration method mentioned in Sect.~\ref{sec:Sel}.
\end{enumerate}
All systematic uncertainties are listed in Table~\ref{tab:SummarySys} and are added in quadrature to give the total systematic uncertainty as they are considered uncorrelated.
\begin{table}[tb!]
\caption[Systematic uncertainties]{Systematic uncertainties in the cross-section calculation for $\sigma_{\decay{\Z}{\mumu}}$. The uncertainties on $\rho$ and $\varepsilon^\text{cand}$ are assumed to be fully correlated between the forward and the backward sample.}
\begin{center}
\begin{tabularx}{\textwidth}{p{6cm}r@{}l@{}lr@{}l@{}l}
\toprule
\textbf{Source} & \multicolumn{3}{p{4.5cm}}{\textbf{Forward}} & \multicolumn{3}{p{4.5cm}}{\textbf{Backward}}\\
\midrule
Sample purity & 0&.5&\,\% & 0&.5&\,\% \\
GEC efficiency &0&.0&\,\% & 1&.9&\,\% \\
Candidate efficiency &8&.4&\,\% & 8&.7&\,\% \\
Multiplicity reweighting &1&.5&\,\% & 2&.0&\,\% \\
Luminosity & 1&.9&\,\% & 2&.1&\,\%\\
\midrule
Total & 8&.8&\,\% & 9&.4&\,\%\\
\bottomrule
\end{tabularx}
\end{center}
\label{tab:SummarySys}
\end{table}
\section{Results\label{sec:Results}}
\begin{figure}[tb!]
\begin{center}
\begin{overpic}[width=0.8\textwidth,scale=.25,tics=20]{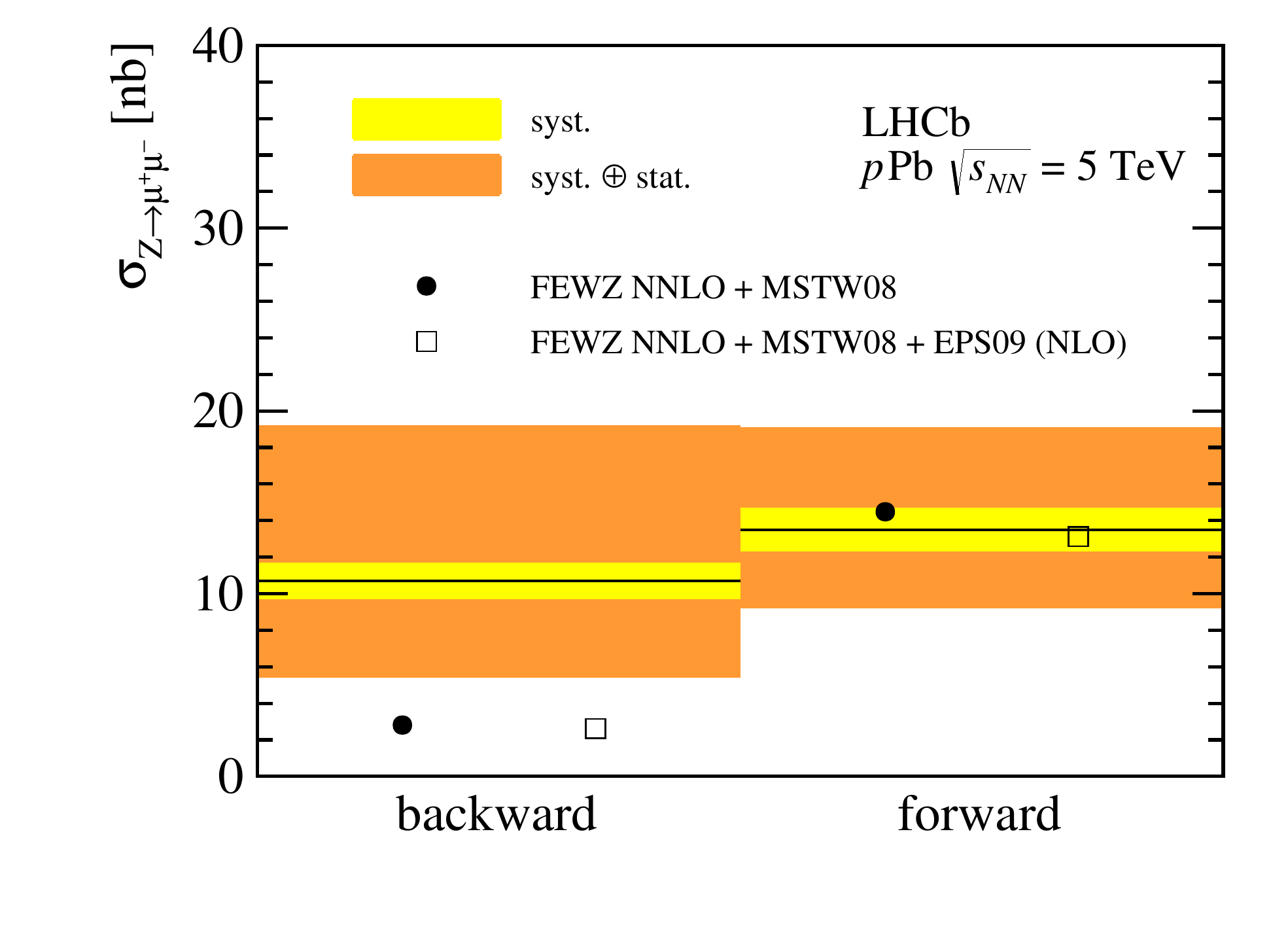}
\end{overpic}
\caption[Cross-section summary]{Experimental results and the theoretical predictions for the \decay{\Z}{\mumu} production cross-section. The inner error bars of the experimental results show the systematic uncertainties. The uncertainties on the theoretical predictions are negligible compared to those on the experimental results.}
\label{fig:Summary}
\end{center}
\end{figure}

The \Z production cross-sections in proton-lead collisions measured in the fiducial region of $60<m_{\mumu}<120\gevcc$, $\pt(\mu^\pm)>20\gevc$ and $2.0<\eta(\mu^\pm)<4.5$ are
\begin{align*}
	\sigma_{\Z\to\mumu}(\text{fwd}) &=\;13.5^{+5.4}_{-4.0}\text{(stat.)}\pm1.2\text{(syst.)}~\text{nb}\\
	\intertext{in the forward direction, and}
	\sigma_{\Z\to\mumu}(\text{bwd}) & =\;10.7^{+8.4}_{-5.1}\text{(stat.)}\pm1.0\text{(syst.)}~\text{nb}
\end{align*}
in the backward direction. The first uncertainty is statistical, defined as the 68\,\% confidence interval with symmetric coverage assuming that the number of candidates follows a Poisson distribution, and the second uncertainty is systematic.\\
\indent The cross-sections calculated at NNLO using \textsc{Fewz}~\cite{Gavin:2011,*Li:2012} and the MSTW08 PDF set~\cite{MSTW08LO} are listed in Table~\ref{tab:NNLO}. The nuclear modifications to the cross-sections are parametrised by the EPS09 nPDF set at next-to-leading order (NLO)~\cite{Eskola:2009}. The obtained values are in agreement with the results of the measurements. The measurement is not precise enough to make any conclusion about the presence of nuclear effects. Figure~\ref{fig:Summary} shows a comparison of the experimental results and the theoretical predictions.\\
\indent The statistical significance of the \Z signal is evaluated by the probability that a Poisson distribution with the expected background yield $N_\text{bkg}^\text{th} = \sigma_{\Z\to\mumu}^{\text{th}}\varepsilon^\text{GEC}\varepsilon^\text{cand}\mathcal{L}(1-\rho)/\rho$ as expectation value fluctuates to the observed signal. The theoretical cross-section $\sigma_{\Z\to\mumu}^{\text{th}}$ is defined as the value obtained from NNLO calculation using \textsc{Fewz} and nuclear modifications based on the EPS09 nPDF set. This gives a significance of $10.4\sigma$ for the \Z signal in the forward direction and $6.8\sigma$ for the backward direction.\\
\begin{table}[b]
\caption[Cross-section predictions]{\decay{\Z}{\mumu} cross-section predictions based on NNLO calculations using \textsc{Fewz} with the MSTW08 PDF set. The nuclear modifications are parametrised by the EPS09 nPDF set. The first uncertainty is the scale uncertainty evaluated by changing the renormalisation and factorisation scales by a factor two up and down. The second one comes from the uncertainties on the PDF and the third in case of the predictions with nuclear effects from the nPDF.}
\begin{center}
\begin{tabularx}{\textwidth}{p{6cm}r@{}l@{}l@{}lr@{}l@{}l@{}l}
\toprule
\textbf{Theory setup} &\multicolumn{7}{c}{$\begin{boldmath}\sigma_{\decay{\Z}{\mumu}}\end{boldmath}$ \textbf{[nb]}}\\
& \multicolumn{4}{p{5cm}}{\textbf{Forward}} & \multicolumn{4}{p{5cm}}{\textbf{Backward}}\\
\midrule
No nuclear effects & $14.48$ & $^{+0.12}_{-0.09}$ & $^{+0.30}_{-0.27}$ && $2.81$ & $^{+0.03}_{-0.03}$ & $^{+0.07}_{-0.06}$\\[0.2cm]
With nuclear effects (EPS09) & $13.12$ & $^{+0.11}_{-0.08}$ & $^{+0.27}_{-0.24}$ & $^{+0.03}_{-0.10}$ & $2.61$ & $^{+0.03}_{-0.03}$ & $^{+0.07}_{-0.06}$ & $^{+0.03}_{-0.08}$\\
\bottomrule
\end{tabularx}
\end{center}
\label{tab:NNLO}
\end{table}
\indent As shown by the calculations using nPDF sets, but also from theoretical calculations~\cite{Guzey:2013} using the leading-twist approach to model the effect of shadowing and anti-shadowing~\cite{Frankfurt:2012}, a suppression of the cross-section in the forward direction is expected while for the backward direction only a small effect is predicted.\\
\indent A particularly sensitive variable to detect nuclear modifications is the forward-backward ratio, $R_\text{FB}$, defined as
\begin{equation}
R_\text{FB}(\sqrt{s_{NN}},|y|) \equiv \frac{\sigma(\sqrt{s_{NN}},+|y|)}{\sigma(\sqrt{s_{NN}},-|y|)}.
\end{equation}
\indent In LHCb, $R_\text{FB}$ can be measured in the overlap region $2.5<|y|<4.0$ of the \Z rapidity in the centre-of-mass frame of the two beam configurations. The cross-section ratio determined for this range including acceptance corrections should be one without any nuclear effects. The predicted value from NNLO calculation using \textsc{Fewz} with the MSTW08 PDF set and the nuclear modifications of the PDF from the EPS09 set at NLO is $R_\text{FB}(2.5<|y|<4.0) = 0.943^{+0.012}_{-0.027}$; the quoted uncertainty is due to the uncertainties on the nPDF. The scale and PDF uncertainties cancel in the ratio. \\
\indent The value of $R_\text{FB}$ measured in the overlap region $2.5<|y|<4.0$ is defined as
\begin{equation}
R_\text{FB}(2.5<|y|<4.0) = \frac{N_\text{cand,fwd}}{N_\text{cand,bwd}}\times\frac{\rho_\text{fwd}}{\rho_\text{bwd}}\times\frac{\varepsilon_\text{bwd}}{\varepsilon_\text{fwd}}\times\frac{\mathcal{L}_\text{bwd}}{\mathcal{L}_\text{fwd}}\times\frac{1}{\beta},
\end{equation}
where $\beta$ is the correction factor for the difference in the detector acceptance of the muons between the forward and backward directions. It is evaluated using NNLO \textsc{Fewz} calculations to be $\beta = 2.419^{+0.127}_{-0.000}\text{(theo.)}\pm0.008\text{(num.)}_{-0.010}^{+0.009}\text{(PDF)}$, where the first uncertainty is from the variation of the renormalisation and factorisation scale, the second the numerical and the last the uncertainty from the PDF uncertainties. The scale variation always leads to an enhancement of $\beta$.\\
\indent The numbers of candidates in the common $y$ range are 2 in the forward and 4 in the backward samples. The measured value for $R_\text{FB}$ is
\begin{equation*}
R_\text{FB}(2.5<|y|<4.0) =  0.094^{+0.104}_{-0.062}\text{(stat.)}_{-0.007}^{+0.004}\text{(syst.)},
\end{equation*}
where the first uncertainty is statistical, defined as the 68\,\% confidence interval with symmetric coverage. The 99.7\,\% (\ie $3\sigma$) confidence interval with symmetric coverage is $[0.002, 1.626]$ whereas the asymmetry of the interval around the central value is due to non-Gaussian statistical uncertainties. The second uncertainty is systematic and includes also the uncertainty on the acceptance correction factor $\beta$. The systematic uncertainties between the forward and the backward directions on the purity and the reconstruction, selection, trigger and muon-identification efficiency are assumed to be fully correlated. The probability to observe a value of $R_\text{FB}$ no larger than that measured, assuming no nuclear modifications (\ie the true value is $R_\text{FB}=1$), is 1.2\;\%. This corresponds to a deviation with a $2.2\sigma$ significance. The probability is estimated with a toy Monte Carlo assuming Poissonian distributions for the number of candidates.
\section{Conclusions\label{sec:Concl}}
Measurements of \Z production in proton-lead collisions at LHCb at $\sqrt{s_{NN}}=5\tev$ in the forward and backward directions, with data corresponding to 1.1 and 0.5~nb$^{-1}$ of integrated luminosity, respectively, have been performed.\\
\indent The cross-section is measured to be
\begin{align*}
	\sigma_{\Z\to\mumu}(\text{fwd})&=\;13.5^{+5.4}_{-4.0}\text{(stat.)}\pm1.2\text{(syst.)}~\text{nb},\\
	\intertext{in the forward direction, and}
	\sigma_{\Z\to\mumu}(\text{bwd}) & =\;10.7^{+8.4}_{-5.1}\text{(stat.)}\pm1.0\text{(syst.)}~\text{nb}
\end{align*}
in the backward direction. These values are in agreement with predictions, although the production of $\Z$ bosons in the backward direction appears to be higher than predictions. The fiducial region is defined by $2.0<\eta(\mu^\pm)<4.5$, $\pt(\mu^\pm)>20\gevc$ and $60<m_{\mumu}<120\gevcc$. The forward-backward ratio $R_\text{FB}$ in the \Z rapidity interval $2.5<|y|<4.0$ in the centre-of-mass frame is found to be lower than expectations, and corresponds to a $2.2\sigma$ deviation from $R_\text{FB}=1$.\\
\indent The statistical precision of the measured cross-sections prevents conclusions on the presence of nuclear effects. The first observation of \Z production in proton-nucleus collisions demonstrates the excellent potential of the study of electroweak bosons in proton-lead collisions at LHCb.

\section*{Acknowledgements}
\noindent We express our gratitude to our colleagues in the CERN
accelerator departments for the excellent performance of the LHC. We
thank the technical and administrative staff at the LHCb
institutes. We acknowledge support from CERN and from the national
agencies: CAPES, CNPq, FAPERJ and FINEP (Brazil); NSFC (China);
CNRS/IN2P3 (France); BMBF, DFG, HGF and MPG (Germany); SFI (Ireland); INFN (Italy);
FOM and NWO (The Netherlands); MNiSW and NCN (Poland); MEN/IFA (Romania);
MinES and FANO (Russia); MinECo (Spain); SNSF and SER (Switzerland);
NASU (Ukraine); STFC (United Kingdom); NSF (USA).
The Tier1 computing centres are supported by IN2P3 (France), KIT and BMBF
(Germany), INFN (Italy), NWO and SURF (The Netherlands), PIC (Spain), GridPP
(United Kingdom).
We are indebted to the communities behind the multiple open
source software packages on which we depend. We are also thankful for the
computing resources and the access to software R\&D tools provided by Yandex LLC (Russia).
Individual groups or members have received support from
EPLANET, Marie Sk\l{}odowska-Curie Actions and ERC (European Union),
Conseil g\'{e}n\'{e}ral de Haute-Savoie, Labex ENIGMASS and OCEVU,
R\'{e}gion Auvergne (France), RFBR (Russia), XuntaGal and GENCAT (Spain), Royal Society and Royal
Commission for the Exhibition of 1851 (United Kingdom).

\clearpage

\addcontentsline{toc}{section}{References}
\setboolean{inbibliography}{true}
\bibliographystyle{LHCb}
\bibliography{main,LHCb-PAPER,LHCb-CONF,LHCb-DP,Library}
\clearpage
\end{document}